\documentclass[prl,nofootinbib,floatfix,twocolumn,showpacs,superscriptaddress]{revtex4}
\usepackage[dvips]{pstcol}
\usepackage{graphicx,amsmath}
\usepackage{dcolumn}
\usepackage{bm}
\usepackage{rotating}
\usepackage{amsmath,amssymb,amsfonts}
\usepackage{color}	
\usepackage{multirow}

\def\PPP{\csname PPP\endcsname}
\expandafter\def\csname PPP\endcsname{\thetable}

\newcommand{\be}{\begin{equation}}
\newcommand{\ee}{\end{equation}}
\newcommand{\beqa}{\begin{eqnarray}}
\newcommand{\eeqa}{\end{eqnarray}}
\newcommand{\nn}{\nonumber}

\definecolor{lightred}{rgb}{1,0.75,0.75}
\definecolor{lightgreen}{rgb}{0.8,1,0.8}

\graphicspath{{./figs/}{./xfigs/}}

\begin{document}

\preprint{DESY/??-???}
\title{Search for a Two-Photon Exchange Contribution to \\
 Inclusive  Deep-Inelastic Scattering  \\
}

\def\groupargonne{\affiliation{Physics Division, Argonne National Laboratory, Argonne, Illinois 60439-4843, USA}}
\def\groupbari{\affiliation{Istituto Nazionale di Fisica Nucleare, Sezione di Bari, 70124 Bari, Italy}}
\def\groupbeijing{\affiliation{School of Physics, Peking University, Beijing 100871, China}}
\def\groupcolorado{\affiliation{Nuclear Physics Laboratory, University of Colorado, Boulder, Colorado 80309-0390, USA}}
\def\groupdesy{\affiliation{{\sc Desy}, 22603 Hamburg, Germany}}
\def\groupzeuthen{\affiliation{{\sc Desy}, 15738 Zeuthen, Germany}}
\def\groupdubna{\affiliation{Joint Institute for Nuclear Research, 141980 Dubna, Russia}}
\def\grouperlangen{\affiliation{Physikalisches Institut, Universit\"at Erlangen-N\"urnberg, 91058 Erlangen, Germany}}
\def\groupferrara{\affiliation{Istituto Nazionale di Fisica Nucleare, Sezione di Ferrara and Dipartimento di Fisica, Universit\`a di Ferrara, 44100 Ferrara, Italy}}
\def\groupfrascati{\affiliation{Istituto Nazionale di Fisica Nucleare, Laboratori Nazionali di Frascati, 00044 Frascati, Italy}}
\def\groupgent{\affiliation{Department of Subatomic and Radiation Physics, University of Gent, 9000 Gent, Belgium}}
\def\groupgiessen{\affiliation{Physikalisches Institut, Universit\"at Gie{\ss}en, 35392 Gie{\ss}en, Germany}}
\def\groupglasgow{\affiliation{Department of Physics and Astronomy, University of Glasgow, Glasgow G12 8QQ, United Kingdom}}
\def\groupillinois{\affiliation{Department of Physics, University of Illinois, Urbana, Illinois 61801-3080, USA}}
\def\groupmichigan{\affiliation{Randall Laboratory of Physics, University of Michigan, Ann Arbor, Michigan 48109-1040, USA }}
\def\groupmoscow{\affiliation{Lebedev Physical Institute, 117924 Moscow, Russia}}
\def\groupnikhef{\affiliation{National Institute for Subatomic Physics (Nikhef), 1009 DB Amsterdam, The Netherlands}}
\def\groupstpetersburg{\affiliation{Petersburg Nuclear Physics Institute, Gatchina, Leningrad region 188300, Russia}}
\def\groupprotvino{\affiliation{Institute for High Energy Physics, Protvino, Moscow region 142281, Russia}}
\def\groupregensburg{\affiliation{Institut f\"ur Theoretische Physik, Universit\"at Regensburg, 93040 Regensburg, Germany}}
\def\grouprome{\affiliation{Istituto Nazionale di Fisica Nucleare, Sezione Roma 1, Gruppo Sanit\`a and Physics Laboratory, Istituto Superiore di Sanit\`a, 00161 Roma, Italy}}
\def\grouptriumf{\affiliation{{\sc Triumf}, Vancouver, British Columbia V6T 2A3, Canada}}
\def\grouptokyo{\affiliation{Department of Physics, Tokyo Institute of Technology, Tokyo 152, Japan}}
\def\groupamsterdam{\affiliation{Department of Physics, VU University, 1081 HV Amsterdam, The Netherlands}}
\def\groupwarsaw{\affiliation{Andrzej Soltan Institute for Nuclear Studies, 00-689 Warsaw, Poland}}
\def\groupyerevan{\affiliation{Yerevan Physics Institute, 375036 Yerevan, Armenia}}
\def\groupnone{\noaffiliation}


\groupargonne
\groupbari
\groupbeijing
\groupcolorado
\groupdesy
\groupzeuthen
\groupdubna
\grouperlangen
\groupferrara
\groupfrascati
\groupgent
\groupgiessen
\groupglasgow
\groupillinois
\groupmichigan
\groupmoscow
\groupnikhef
\groupstpetersburg
\groupprotvino
\groupregensburg
\grouprome
\grouptriumf
\grouptokyo
\groupamsterdam
\groupwarsaw
\groupyerevan


\author{A.~Airapetian}  \groupgiessen \groupmichigan
\author{N.~Akopov}  \groupyerevan
\author{Z.~Akopov}  \groupdesy
\author{E.C.~Aschenauer}  \groupzeuthen
\author{W.~Augustyniak}  \groupwarsaw
\author{R.~Avakian} \groupyerevan
\author{A.~Avetissian}  \groupyerevan
\author{E.~Avetisyan}  \groupdesy
\author{B.~Ball}  \groupmichigan
\author{S.~Belostotski}  \groupstpetersburg
% ...

\collaboration{The {\sc Hermes} Collaboration} \noaffiliation

\date{\today}

\date{\today}

\begin{abstract}

The transverse-target single-spin asymmetry for inclusive deep-inelastic scattering 
with effectively unpolarized electron and positron 
beams off a  transversely polarized hydrogen target was measured,
with the goal of searching for a two-photon exchange 
signal in the kinematic range $0.007 < x_B < 0.9$ and 
0.25~GeV$^2 < Q^2 <$~20~GeV$^2$.
In two separate regions $Q^2>$~1~GeV$^2$ and $Q^2<$~1~GeV$^2$, and
for both electron and positron beams, the asymmetries are found to be 
consistent with zero
within statistical and systematic uncertainties, which are of order $10^{-3}$ for the asymmetries integrated over $x_B$.
\end{abstract}


\maketitle

In recent years, the contribution of two-photon exchange to the cross section 
for  electron-nucleon scattering has received considerable attention. 
In elastic $ep$ scattering, 
two-photon exchange effects are believed to be the best candidate 
to explain the discrepancy in the measurement of the ratio $G_E/G_M$
of the electric and magnetic form factors of the proton obtained at large
 four-momentum transfer between the Rosenbluth method 
 and the polarization transfer method~\cite{hyde}. 
It has been shown
that the interference between the one-photon
 and two-photon exchange amplitudes can 
 affect the Rosenbluth extraction of the nucleon form factors
at the level of a few percent. This is  enough  to explain most of the discrepancy
between the results of the two  methods~\cite{elgv,bm}, although none of the recent 
calculations can fully resolve the discrepancy at all momentum transfers~\cite{aw}.
Two-photon exchange effects have also been shown
to affect
the measurement of parity violation in elastic  scattering of
 longitudinally polarized electrons off 
unpolarized protons,
with corrections of several percent to the parity-violating asymmetry~\cite{AC2005}.

In order to investigate contributions from  two-photon exchange,
it is necessary to find experimental observables that allow their isolation.
Beam-charge and transverse single-spin asymmetries (SSAs) are  two  suitable 
candidates.
In both elastic and inclusive inelastic lepton-nucleon scattering,
these asymmetries arise from the interference of one-photon and 
two-photon exchange amplitudes.
Specifically, beam-charge asymmetries in the unpolarized cross section 
arise from the real part of the two-photon exchange amplitude~\cite{mar},
while inclusive  transverse SSAs are sensitive to the imaginary
part~\cite{Metz2006}.

To date, all evidence of non-zero two-photon exchange effects  in lepton-nucleon 
interactions comes
from elastic scattering, $l + N \rightarrow l' + N'$.
Measurements of the cross-section ratio $R = \sigma_{e^+p} / \sigma_{e^-p}$ 
are compiled in Ref.~\cite{mar}. Though the individual measurements are
consistent with $R$ being unity, 
a recent reanalysis~\cite{arrington} demonstrates
that a deviation of about 5\% at low values of 
four-momentum transfer  and  virtual-photon 
polarization is not excluded. 
 Three experiments have measured a non-zero transverse-beam SSA 
of order $10^{-5}-10^{-6}$
in elastic scattering of transversely polarized electrons
off unpolarized protons~\cite{wells,maas,armstrong}.

In inelastic scattering
  no clear signature of two-photon exchange effects
has yet been observed. Measurements of the cross-section ratio $R$
with $e^+/e^-$ and $\mu^+/\mu^-$ 
beams~\cite{jostlein,hartwig,fancher1,rochester,hartwig2,aubert,argento}
show no effect within their accuracy of a few percent.
The transverse-target SSA has  been measured 
at the Cambridge Electron
Accelerator~\cite{appel,cambridge} and at {\sc Slac}~\cite{slac35}.
The data are confined to the region of nucleon resonances, and 
show an asymmetry which is compatible with zero within the few-percent level
of the experimental uncertainties.

In inclusive 
 deep-inelastic scattering (DIS),  $l +p \rightarrow l'+X$,
and in the one-photon exchange approximation, 
such a SSA is forbidden by the combination 
of time reversal invariance, 
parity conservation, and the hermiticity of the
electromagnetic current operator, as stated in the Christ-Lee
theorem~\cite{christ-lee}.
A non-zero SSA can therefore be interpreted as an indication
of two-photon exchange. 

Ref.~\cite{Metz2006} presents a theoretical treatment of the 
transverse SSA arising from the 
interference of one-photon and two-photon exchange amplitudes in DIS. 
For an unpolarized beam (U) and a transversely (T) polarized nucleon target, 
the spin-dependent part of the  cross section is  given by
\begin{equation}
\label{sigma}
\sigma_{UT} \propto  e_l\alpha_{em}~\frac{M}{Q}~\varepsilon_{\mu \nu \rho \sigma}~ S^{\mu}p^{\nu}k^{\rho} k'^\sigma
	~C_T.
\end{equation}
Here, $e_l$ is the charge of the incident lepton, 
$M$ is the 
nucleon mass, $-Q^2$ is the squared four-momentum transfer, 
$p$, $k$ and $k'$ are  the four-momenta
of the target, the incident and the scattered lepton, respectively, 
while $\varepsilon_{\mu \nu \rho \sigma} $ is the Levi-Civita tensor.
The term $\varepsilon_{\mu \nu \rho \sigma} 
S^{\mu}p^{\nu}k^{\rho} k'^\sigma$ 
is proportional to $\vec{S}\cdot (\vec{k}\times \vec{k'})$, consequently the largest 
asymmetry  is obtained when  the spin vector $\vec{S}$ 
  is perpendicular to the 
 lepton scattering plane defined by the three-momenta $\vec{k}$ and
 $\vec{k'}$.
 Finally, $C_T$ is a higher-twist term arising from quark-quark and quark-gluon-quark
 correlations.

As $\sigma_{UT}$  is proportional to the electromagnetic coupling constant
$\alpha_{em}$, it is expected to be small. 
Furthermore, due to the factor $M/Q$ in Eq.~(\ref{sigma}), $\sigma_{UT}$ is expected to increase with 
decreasing $Q^2$.
A calculation based on certain model assumptions~\cite{ASW2008}
for a {\sc Jlab} experiment~\cite{jlab-07013} yields
expectations for the asymmetry of order $10^{-4}$ at
the kinematics of that experiment.
The authors in Ref.~\cite{Metz2006}, on the other hand, do not exclude asymmetries 
as large as $10^{-2}$ and point out that the term $C_T$ in Eq.~(\ref{sigma}) cannot be 
completely evaluated at present.
Due to the factor  $e_l$ in Eq.~(\ref{sigma}), the asymmetry is expected to have a different sign for 
opposite beam charges.
The capability of the {\sc Hera} accelerator to supply both electron and positron 
beams thus provides an additional means to isolate a possible effect 
from two-photon exchange.

In this paper a first precise measurement  of the transverse-target  SSA
in inclusive DIS of unpolarized electrons and positrons off a
transversely polarized hydrogen target is presented.

The data 
 were collected with the {\sc Hermes} spectrometer~\cite{hermes_spectr} during the period 
2002-2005.
The 27.6~GeV positron or electron 
beam was scattered off the transversely polarized gaseous  hydrogen target internal to the 
{\sc Hera} storage ring at {\sc Desy}.
The open-ended target cell was fed by an atomic-beam source~\cite{21} based on Stern-Gerlach 
separation combined with radio-frequency transitions of hydrogen hyperfine states.
The direction of the target spin vector was reversed  at 1-3~minute time intervals to
minimize systematic effects, while both 
the nuclear polarization and the atomic fraction 
of the target gas inside the storage cell
were continuously measured~\cite{22}.
Data were collected with 
the target polarized transversely to the beam direction,
in both ``upward'' and ``downward'' directions in the laboratory frame.  
The beam was longitudinally polarized, but a helicity-balanced data
sample was used to obtain an effectively unpolarized beam.
Only the scattered leptons were considered in this analysis.
Leptons were distinguished from hadrons by using a transition-radiation
detector, a scintillator pre-shower counter, a dual-radiator ring-imaging Cherenkov detector,
 and an electromagnetic calorimeter.
In order to exclude any contamination from a transverse hadron SSA in the                      
lepton signal, hadrons were suppressed by very stringent particle                               
identification requirements such that their contamination in the lepton                         
sample is smaller than $2\times 10^{-4}$.
This resulted in a lepton identification efficiency greater than 94\%.
Events were selected  in the
kinematic region $0.007<x_B<0.9$, $0.1<y<0.85$, 0.25~GeV$^2<Q^2<20$~GeV$^2$,
and $W^2>4$~GeV$^2$.   Here, $x_B$ is the Bjorken scaling variable, 
$y$ is the fractional beam energy carried by the virtual photon in the laboratory frame, and $W$ is the
 invariant mass of the photon-nucleon system.

The differential yield
for a given target spin direction ($\uparrow$ upwards or $\downarrow$
downwards) can be expressed as
\begin{eqnarray}\label{eq:yields}
&\!&\!\!\!\!\! \frac{\mathrm{d^3}N^{\uparrow(\downarrow)}}{\mathrm{d}x_B~\mathrm{d}Q^2~\mathrm{d}\phi_S} =  \nn\\
& &\left[ L^{\uparrow(\downarrow)}~\mathrm{d^3}\sigma_{UU} +(-)
 L_P^{\uparrow(\downarrow)}~\mathrm{d^3}\sigma_{UT}\right]~\Omega(x_B,Q^2,\phi_S) \nn\\ 
&\!&\!\!\!\!\!=\, \mathrm{d^3}\sigma_{UU}
\left[
 L^{\uparrow(\downarrow)} +(-)\right. \nn\\
&\!&\!\!\!\!\!
\left. \,\,\,
 L_P^{\uparrow(\downarrow)}~A_{UT}^{\sin\phi_S}(x_B,Q^2) \sin\phi_S\right]~
\Omega(x_B,Q^2,\phi_S).
\end{eqnarray}
Here, 
 $\phi_S$ is the azimuthal angle about the beam direction
between the lepton scattering plane and the ``upwards'' target spin direction,
$\sigma_{UU}$ is the unpolarized cross section.
Also,
$L^{\uparrow(\downarrow)}$ is the total luminosity in
the $\uparrow$  ($\downarrow$) polarization state, 
$L_P^{\uparrow(\downarrow)}=\int L^{\uparrow(\downarrow)}(t)~P(t)~\text{d}t$ is
the integrated luminosity weighted by the magnitude $P$ of the target polarization, and  $\Omega$ 
is the detector acceptance efficiency.
The $\sin\phi_S$ azimuthal 
dependence follows directly from the form 
$\vec{S}\cdot (\vec{k}\times \vec{k'})$ 
of the spin-dependent part of the cross section; $A_{UT}^{\sin\phi_S}$ refers to its amplitude.

The asymmetry was calculated as
\begin{equation}\label{asy_def}
A_{UT}(x_B,Q^2,\phi_S)=
\frac{
\displaystyle\frac{N^\uparrow}{L^\uparrow_P}
-      \frac{N^\downarrow}{L^\downarrow_P}
}
{
{\displaystyle\frac{N^\uparrow}{L^\uparrow}
      +\frac{N^\downarrow}{L^\downarrow}}}~,
\end{equation}
where $N^{\uparrow(\downarrow)}$ are the number of events measured in bins of $x_B,~Q^2$, and $\phi_S$.
With the use of Eq.~(\ref{eq:yields}),  it can be approximated, for small 
differences of the two average target polarizations 
$\langle P^{\uparrow(\downarrow)}\rangle =L^{\uparrow(\downarrow)}_P/L^{\uparrow(\downarrow)}$, as
\begin{align}\label{eq4}
A_{UT}(x_B,Q^2,\phi_S)&\simeq 
A_{UT}^{\sin\phi_S}
\sin \phi_S 
+\frac{1}{2}\frac{\langle P^\downarrow \rangle - \langle P^\uparrow \rangle}{\langle P^\uparrow \rangle
\langle  P^\downarrow \rangle}.
\end{align}
As shown in Table~\ref{tab:pol}, $\langle P^\uparrow\rangle $ and $\langle P^\downarrow\rangle$ 
are the same to a good 
approximation for all data-taking periods.
\begin{table}[!t]
\begin{ruledtabular}
\begin{tabular}{|c|c|c|c|c|}
~year~&~beam~&~$\langle P^\uparrow\rangle$~&~$\langle P^\downarrow\rangle $~
&~Events~\\
\hline
 2002 & $e^+$&0.783$\pm$0.041 & 0.783$\pm$0.041& 0.9~M\\
 2004 & $e^+$&0.745$\pm$0.054 & 0.742$\pm$0.054& 2.0~M\\
 2005 & $e^-$&0.705$\pm$0.065 & 0.705$\pm$0.065& 4.8~M\\
\end{tabular}
\end{ruledtabular}
\caption{\label{tab:pol} Average target polarizations  and total
  number of inclusive events for the three data sets
used in this analysis. }
\end{table}

The advantage of using the fully-differential asymmetry 
$A_{UT}(x_B,Q^2,\phi_S)$ in Eq.~(\ref{asy_def}) instead of the more
common left-right asymmetry $A_{N}(x_B,Q^2)$ is that the acceptance function 
$\Omega$ cancels in each $(x_B,Q^2,\phi_S)$ kinematic bin, 
if the bin size or the asymmetry is small. 
Assuming the $\phi_S$ dependence of $\sigma_{UT}$ in Eq.~(\ref{sigma}) and Eq.~(\ref{eq:yields}),
it can be easily shown that the $\sin\phi_S$ amplitude $A_{UT}^{\sin\phi_S}$ 
and the  left-right normal asymmetry $A_{N}$ are related by
\begin{eqnarray}
&\!&\!\!\!\!\!  A_{N} =   \frac{\sigma_L -\sigma_R}{\sigma_L + \sigma_R} = \nn \\  
& & \frac{\int_0^{\pi}d\phi_S~\mathrm{d^3}\sigma_{UU}~A_{UT}^{\sin\phi_S}~\sin\phi_S}{\int_0^{\pi}d
\phi_S~\mathrm{d^3}\sigma_{UU}}=\frac{2}{\pi}~A_{UT}^{\sin\phi_S},
\label{eq:4pi}
\end{eqnarray}
 where $\sigma_L$~($\sigma_R$) refers to the integrated cross section within the angular range $0\le\phi_S<\pi$~($\pi\le\phi_S<2\pi$).

For this analysis the $Q^2$ range was divided into a ``DIS region''
with $Q^2>$~1~GeV$^2$ 
and a ``low-$Q^2$ region'' with $Q^2<$~1~GeV$^2$. 
To test for a possible enhancement of the transverse-target SSA due to the factor $M/Q$
appearing in Eq.~(\ref{sigma}) the data at low $Q^2$ are also presented,
 though, strictly speaking,
Eq.~(\ref{sigma}) may not be applicable to this range.

The $A_{UT}^{\sin\phi_S}$ amplitudes were extracted with a binned $\chi^2$ 
fit of the functional form $p_1 \sin\phi_S+p_2$ to the measured asymmetry. 
Leaving $p_2$ as a free parameter or fixing it to the values given by 
Eq.~(\ref{eq4}) and Table~\ref{tab:pol}
had no impact on the extracted $\sin\phi_S$ amplitude $p_1\equiv A_{UT}^{\sin\phi_S}$.

\begin{figure}[ht]
\hspace*{-0.0cm}
\includegraphics[width=1.0\columnwidth]{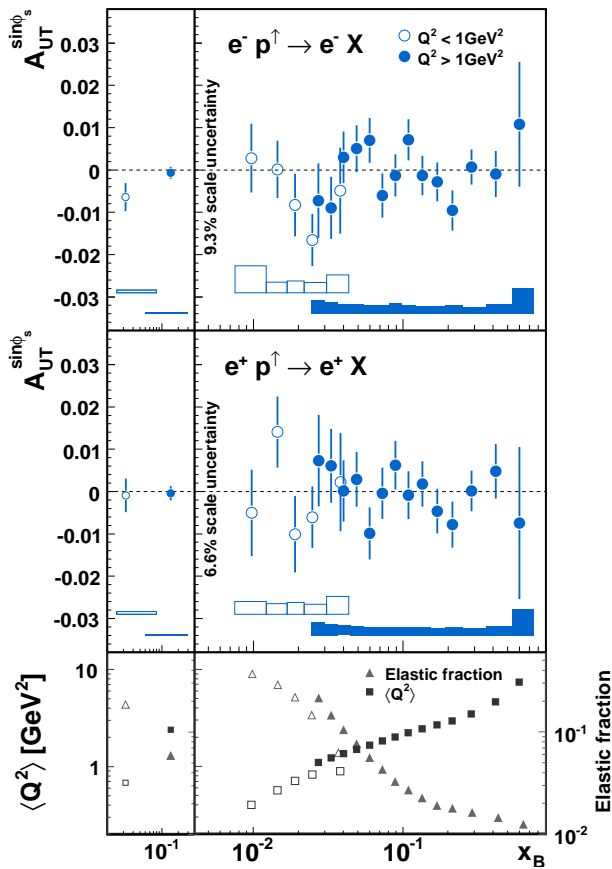}
\caption{\label{fig:info} The $x_B$ dependence of the 
$\sin \phi_S$ amplitudes  $A_{UT}^{\sin\phi_S}$
  measured with an electron beam (top) and a positron beam (center). 
  The open (closed) circles identify the data 
  with $Q^2<1$ GeV$^2$ ($Q^2>1$ GeV$^2$). The error bars show the statistical 
  uncertainties, while the error boxes show the systematic uncertainties.
The asymmetries integrated over $x_B$ are shown on the left.
  Bottom panel: average $Q^2$ vs.~$x_B$ from data (squares), and the
  fraction of elastic background events to the total event sample from a Monte Carlo simulation (triangles).}
\end{figure}

The final results for the measured $\sin \phi_S$ amplitudes  $A^{\sin\phi_S}_{UT}$
are shown in
Fig.~\ref{fig:info} as a function of $x_B$ separately for electrons and positrons.
 In both cases the asymmetries are consistent with zero within their
uncertainties.
Due to the kinematics of the experiment,
the quantities $x_B$ and $\langle Q^2\rangle $ are strongly correlated, as
shown in the bottom panel of Fig.~\ref{fig:info}.

The resulting amplitudes were not corrected for kinematic migration of                         
inelastic events due to detector smearing and higher order QED effects or                      
contamination by the radiative tail from elastic scattering.
The latter correction requires knowledge of the presently 
unknown elastic two-photon asymmetry.
Instead, the contribution of the elastic radiative tail to 
the total event sample was estimated from a Monte Carlo simulation
based on the {\sc Lepto} generator~\cite{lep}
 together with the {\sc Radgen}~\cite{rad}
 determination 
of QED radiative effects and with a {\sc Geant}~\cite{gea} 
based simulation of the detector.
The elastic fraction is shown in the lower panel of Fig.~\ref{fig:info}. It
reaches values as high as about 35\% in the lowest $x_B$ bin, where 
$y$ is large ($\langle y\rangle\simeq 0.80$) and hence radiative corrections are
largest~\cite{MoTsai}. The elastic fraction 
rapidly decreases
towards high $x_B$, becoming less than 3\% for $x_B>0.1$.

The systematic uncertainties, shown in the fourth column of Table II 
and as error boxes in Fig.~\ref{fig:info},
include contributions due to corrections for misalignment of the detector,
 beam position and 
slope at the interaction point and bending of the beam and the scattered lepton in the transverse 
holding field of the target magnet.
They were determined from a high statistics Monte Carlo sample obtained from a simulation 
containing a full description of the detector, where an artificial spin-dependent azimuthal 
asymmetry was implemented. 
Input asymmetries being zero or as small as $10^{-3}$ were well reproduced within the statistical 
uncertainty of the Monte Carlo sample, which was about five times smaller than the statistical uncertainty 
of the data. For each measured point the systematic uncertainty was obtained 
as the maximum value of either 
the statistical uncertainty of the Monte Carlo sample or the difference between the 
input asymmetry and the extracted one. Systematic uncertainties from other sources like particle identification or
 trigger efficiencies were found to be negligible.

The transverse single-spin asymmetry amplitudes
  $A_{UT}^{\sin\phi_S}$ for electron and positron beams
integrated over $x_B$ are given separately
for the ``low-$Q^2$ region'' and the ``DIS region''  in Table~\ref{final-table}
 along with their statistical and systematic uncertainties. 
All asymmetry amplitudes are consistent with zero within
their uncertainties, which in the DIS region are of order $10^{-3}$. The only
exception is the low-$Q^2$ electron sample, where the asymmetry is 
1.9 standard deviations different from zero.
No hint of a sign change between electron and positron asymmetries  
is observed within uncertainties.
\begin{table}[t!]
\begin{ruledtabular}
\begin{tabular}{|c|c|c|c|c|c|}
beam  & $A_{UT}^{\sin\phi_S}$& $\delta A_{UT}^{\sin\phi_S}$(stat.)   
&$\delta A_{UT}^{\sin\phi_S}$(syst.)     &$\langle x_B\rangle $& $\langle Q^2\rangle$\\
     & $\times 10^{-3}$&   $\times 10^{-3}$  & $\times 10^{-3}$&&[GeV$^2$]\\
\hline 
\hline
$e^+$        &  -0.61  & 3.97     & 0.63  & \multirow{2}{*}{0.02} & \multirow{2}{*}{0.68} \\
$e^-$        &  -6.55  & 3.40     & 0.63  &  & \\  
\hline 
$e^+$        &  -0.60  & 1.70     & 0.29  & \multirow{2}{*}{0.14} & \multirow{2}{*}{2.40} \\
$e^-$        &  -0.85  & 1.50     & 0.29  &      & 
\end{tabular}
\end{ruledtabular}
\caption{\label{final-table} 
	The integrated transverse single-spin asymmetry 
	amplitude $A_{UT}^{\sin\phi_S}$  with its statistical and systematic 
	uncertainties and the average values for $x_B$ and $Q^2$ 
	measured separately 
	for electron and positron beams in the two $Q^2$ ranges 
	$Q^2<1$~GeV$^2$ (upper rows) and $Q^2>1$~GeV$^2$ (lower rows). 
The systematic uncertainties
 contain the effects of detector misalignment and beam 
position and slope at the target, as estimated by a Monte Carlo simulation, 
but not the scale uncertainties from the target polarization which amounts 
to 9.3\% (6.6\%) for the electron (positron) sample.
Also, the results are not corrected for smearing, radiative effects and elastic background events.}
\end{table}

In conclusion,  single-spin asymmetries were
measured in inclusive deep-inelastic scattering 
at {\sc Hermes} with unpolarized electron and positron beams and
a transversely polarized hydrogen target with the goal of searching for
a signal of two-photon exchange. No signal was found
within the uncertainties, which are of order $10^{-3}$.

We gratefully acknowledge the {\sc Desy} management for its support and the staff
at {\sc Desy} and the collaborating institutions for their significant effort.
This work was supported by the FWO-Flanders and IWT, Belgium;
the Natural Sciences and Engineering Research Council of Canada;
the National Natural Science Foundation of China;
the Alexander von Humboldt Stiftung;
the German Bundesministerium f\"ur Bildung und Forschung (BMBF);
the Deutsche Forschungsgemeinschaft (DFG);
the Italian Istituto Nazionale di Fisica Nucleare (INFN);
the MEXT, JSPS, and G-COE of Japan;
the Dutch Foundation for Fundamenteel Onderzoek der Materie (FOM);
the U.~K.~Engineering and Physical Sciences Research Council, 
the Science and Technology Facilities Council,
and the Scottish Universities Physics Alliance;
the U.~S.~Department of Energy (DOE) and the National Science Foundation (NSF);
the Russian Academy of Science and the Russian Federal Agency for 
Science and Innovations;
the Ministry of Economy and the Ministry of Education and Science of 
Armenia;
and the European Community-Research Infrastructure Activity under the
FP6 ''Structuring the European Research Area'' program
(HadronPhysics, contract number RII3-CT-2004-506078).

\bibliography{paper}

\begin{thebibliography}{31}
\expandafter\ifx\csname natexlab\endcsname\relax\def\natexlab#1{#1}\fi
\expandafter\ifx\csname bibnamefont\endcsname\relax
  \def\bibnamefont#1{#1}\fi
\expandafter\ifx\csname bibfnamefont\endcsname\relax
  \def\bibfnamefont#1{#1}\fi
\expandafter\ifx\csname citenamefont\endcsname\relax
  \def\citenamefont#1{#1}\fi
\expandafter\ifx\csname url\endcsname\relax
  \def\url#1{\texttt{#1}}\fi
\expandafter\ifx\csname urlprefix\endcsname\relax\def\urlprefix{URL }\fi
\providecommand{\bibinfo}[2]{#2}
\providecommand{\eprint}[2][]{\url{#2}}

\bibitem[{\citenamefont{Hyde and de~Jager}(2004)}]{hyde}
\bibinfo{author}{\bibfnamefont{C.~E.} \bibnamefont{Hyde}} \bibnamefont{and}
  \bibinfo{author}{\bibfnamefont{K.}~\bibnamefont{de~Jager}},
  \bibinfo{journal}{Ann. Rev. Nucl. Part. Sci.} \textbf{\bibinfo{volume}{54}},
  \bibinfo{pages}{217} (\bibinfo{year}{2004}).

\bibitem[{\citenamefont{Guichon and Vanderhaeghen}(2003)}]{elgv}
\bibinfo{author}{\bibfnamefont{P.~A.~M.} \bibnamefont{Guichon}}
  \bibnamefont{and}
  \bibinfo{author}{\bibfnamefont{M.}~\bibnamefont{Vanderhaeghen}},
  \bibinfo{journal}{Phys. Rev. Lett.} \textbf{\bibinfo{volume}{91}},
  \bibinfo{pages}{142303} (\bibinfo{year}{2003}).

\bibitem[{\citenamefont{Blunden et~al.}(2003)\citenamefont{Blunden,
  Melnitchouk, and Tjon}}]{bm}
\bibinfo{author}{\bibfnamefont{P.~G.} \bibnamefont{Blunden}},
  \bibinfo{author}{\bibfnamefont{W.}~\bibnamefont{Melnitchouk}},
  \bibnamefont{and} \bibinfo{author}{\bibfnamefont{J.~A.} \bibnamefont{Tjon}},
  \bibinfo{journal}{Phys. Rev. Lett.} \textbf{\bibinfo{volume}{91}},
  \bibinfo{pages}{142304} (\bibinfo{year}{2003}).

\bibitem[{\citenamefont{Arrington et~al.}(2007)\citenamefont{Arrington,
  Melnitchouk, and Tjon}}]{aw}
\bibinfo{author}{\bibfnamefont{J.}~\bibnamefont{Arrington}},
  \bibinfo{author}{\bibfnamefont{W.}~\bibnamefont{Melnitchouk}},
  \bibnamefont{and} \bibinfo{author}{\bibfnamefont{J.~A.} \bibnamefont{Tjon}},
  \bibinfo{journal}{Phys. Rev.} \textbf{\bibinfo{volume}{C76}},
  \bibinfo{pages}{035205} (\bibinfo{year}{2007}).

\bibitem[{\citenamefont{Afanasev and Carlson}(2005)}]{AC2005}
\bibinfo{author}{\bibfnamefont{A.~V.} \bibnamefont{Afanasev}} \bibnamefont{and}
  \bibinfo{author}{\bibfnamefont{C.~E.} \bibnamefont{Carlson}},
  \bibinfo{journal}{Phys. Rev. Lett.} \textbf{\bibinfo{volume}{94}},
  \bibinfo{pages}{212301} (\bibinfo{year}{2005}).

\bibitem[{\citenamefont{Mar et~al.}(1968)}]{mar}
\bibinfo{author}{\bibfnamefont{J.}~\bibnamefont{Mar}} \bibnamefont{et~al.},
  \bibinfo{journal}{Phys. Rev. Lett.} \textbf{\bibinfo{volume}{21}},
  \bibinfo{pages}{482} (\bibinfo{year}{1968}).

\bibitem[{\citenamefont{Metz et~al.}(2006)\citenamefont{Metz, Schlegel, and
  Goeke}}]{Metz2006}
\bibinfo{author}{\bibfnamefont{A.}~\bibnamefont{Metz}},
  \bibinfo{author}{\bibfnamefont{M.}~\bibnamefont{Schlegel}}, \bibnamefont{and}
  \bibinfo{author}{\bibfnamefont{K.}~\bibnamefont{Goeke}},
  \bibinfo{journal}{Phys. Lett.} \textbf{\bibinfo{volume}{B643}},
  \bibinfo{pages}{319} (\bibinfo{year}{2006}).

\bibitem[{\citenamefont{Arrington}(2004)}]{arrington}
\bibinfo{author}{\bibfnamefont{J.}~\bibnamefont{Arrington}},
  \bibinfo{journal}{Phys. Rev. C} \textbf{\bibinfo{volume}{69}},
  \bibinfo{pages}{032201} (\bibinfo{year}{2004}).

\bibitem[{\citenamefont{Wells et~al.}(2001)}]{wells}
\bibinfo{author}{\bibfnamefont{S.~P.} \bibnamefont{Wells}} \bibnamefont{et~al.}
  (\bibinfo{collaboration}{{\sc Sample} Collaboration}),
  \bibinfo{journal}{Phys. Rev.} \textbf{\bibinfo{volume}{C63}},
  \bibinfo{pages}{064001} (\bibinfo{year}{2001}).

\bibitem[{\citenamefont{Maas et~al.}(2005)}]{maas}
\bibinfo{author}{\bibfnamefont{F.~E.} \bibnamefont{Maas}} \bibnamefont{et~al.}
  (\bibinfo{collaboration}{{\sc A4} Collaboration}), \bibinfo{journal}{Phys.
  Rev. Lett.} \textbf{\bibinfo{volume}{94}}, \bibinfo{pages}{082001}
  (\bibinfo{year}{2005}).

\bibitem[{\citenamefont{Armstrong et~al.}(2007)}]{armstrong}
\bibinfo{author}{\bibfnamefont{D.~S.} \bibnamefont{Armstrong}}
  \bibnamefont{et~al.} (\bibinfo{collaboration}{{\sc G0} Collaboration}),
  \bibinfo{journal}{Phys. Rev. Lett.} \textbf{\bibinfo{volume}{99}},
  \bibinfo{pages}{092301} (\bibinfo{year}{2007}).

\bibitem[{\citenamefont{Jostlein et~al.}(1974)}]{jostlein}
\bibinfo{author}{\bibfnamefont{H.}~\bibnamefont{Jostlein}}
  \bibnamefont{et~al.}, \bibinfo{journal}{Phys. Lett.}
  \textbf{\bibinfo{volume}{B52}}, \bibinfo{pages}{485} (\bibinfo{year}{1974}).

\bibitem[{\citenamefont{Hartwig et~al.}(1976)}]{hartwig}
\bibinfo{author}{\bibfnamefont{S.}~\bibnamefont{Hartwig}} \bibnamefont{et~al.},
  \bibinfo{journal}{Lett. Nuovo Cim.} \textbf{\bibinfo{volume}{15}},
  \bibinfo{pages}{429} (\bibinfo{year}{1976}).

\bibitem[{\citenamefont{Fancher et~al.}(1976)}]{fancher1}
\bibinfo{author}{\bibfnamefont{D.~L.} \bibnamefont{Fancher}}
  \bibnamefont{et~al.}, \bibinfo{journal}{Phys. Rev. Lett.}
  \textbf{\bibinfo{volume}{37}}, \bibinfo{pages}{1323} (\bibinfo{year}{1976}).

\bibitem[{\citenamefont{Rochester et~al.}(1976)}]{rochester}
\bibinfo{author}{\bibfnamefont{L.~S.} \bibnamefont{Rochester}}
  \bibnamefont{et~al.}, \bibinfo{journal}{Phys. Rev. Lett.}
  \textbf{\bibinfo{volume}{36}}, \bibinfo{pages}{1284} (\bibinfo{year}{1976}).

\bibitem[{\citenamefont{Hartwig et~al.}(1979)}]{hartwig2}
\bibinfo{author}{\bibfnamefont{S.}~\bibnamefont{Hartwig}} \bibnamefont{et~al.},
  \bibinfo{journal}{Phys. Lett.} \textbf{\bibinfo{volume}{B82}},
  \bibinfo{pages}{297} (\bibinfo{year}{1979}).

\bibitem[{\citenamefont{Aubert et~al.}(1986)}]{aubert}
\bibinfo{author}{\bibfnamefont{J.~J.} \bibnamefont{Aubert}}
  \bibnamefont{et~al.} (\bibinfo{collaboration}{{\sc Emc}}),
  \bibinfo{journal}{Nucl. Phys.} \textbf{\bibinfo{volume}{B272}},
  \bibinfo{pages}{158} (\bibinfo{year}{1986}).

\bibitem[{\citenamefont{Argento et~al.}(1984)}]{argento}
\bibinfo{author}{\bibfnamefont{A.}~\bibnamefont{Argento}} \bibnamefont{et~al.}
  (\bibinfo{collaboration}{{\sc Bcdms} Collaboration}), \bibinfo{journal}{Phys.
  Lett.} \textbf{\bibinfo{volume}{B140}}, \bibinfo{pages}{142}
  (\bibinfo{year}{1984}).

\bibitem[{\citenamefont{Appel et~al.}(1970)}]{appel}
\bibinfo{author}{\bibfnamefont{J.~A.} \bibnamefont{Appel}}
  \bibnamefont{et~al.}, \bibinfo{journal}{Phys. Rev.}
  \textbf{\bibinfo{volume}{D1}}, \bibinfo{pages}{1285} (\bibinfo{year}{1970}).

\bibitem[{\citenamefont{Chen et~al.}(1968)}]{cambridge}
\bibinfo{author}{\bibfnamefont{J.~R.} \bibnamefont{Chen}} \bibnamefont{et~al.},
  \bibinfo{journal}{Phys. Rev. Lett.} \textbf{\bibinfo{volume}{21}},
  \bibinfo{pages}{1279} (\bibinfo{year}{1968}).

\bibitem[{\citenamefont{Rock et~al.}(1970)}]{slac35}
\bibinfo{author}{\bibfnamefont{S.}~\bibnamefont{Rock}} \bibnamefont{et~al.},
  \bibinfo{journal}{Phys. Rev. Lett.} \textbf{\bibinfo{volume}{24}},
  \bibinfo{pages}{748} (\bibinfo{year}{1970}).

\bibitem[{\citenamefont{Christ and Lee}(1966)}]{christ-lee}
\bibinfo{author}{\bibfnamefont{N.}~\bibnamefont{Christ}} \bibnamefont{and}
  \bibinfo{author}{\bibfnamefont{T.~D.} \bibnamefont{Lee}},
  \bibinfo{journal}{Phys. Rev.} \textbf{\bibinfo{volume}{143}},
  \bibinfo{pages}{1310} (\bibinfo{year}{1966}).

\bibitem[{\citenamefont{Afanasev et~al.}(2008)\citenamefont{Afanasev, Strikman,
  and Weiss}}]{ASW2008}
\bibinfo{author}{\bibfnamefont{A.}~\bibnamefont{Afanasev}},
  \bibinfo{author}{\bibfnamefont{M.}~\bibnamefont{Strikman}}, \bibnamefont{and}
  \bibinfo{author}{\bibfnamefont{C.}~\bibnamefont{Weiss}},
  \bibinfo{journal}{Phys. Rev.} \textbf{\bibinfo{volume}{D77}},
  \bibinfo{pages}{014028} (\bibinfo{year}{2008}).

\bibitem[{\citenamefont{Jiang et~al.}(2007)}]{jlab-07013}
\bibinfo{author}{\bibfnamefont{X.}~\bibnamefont{Jiang}} \bibnamefont{et~al.},
  \bibinfo{journal}{Jefferson Lab Hall A Experiment E-07-013}
  (\bibinfo{year}{2007}).

\bibitem[{\citenamefont{Ackerstaff et~al.}(1998)}]{hermes_spectr}
\bibinfo{author}{\bibfnamefont{K.}~\bibnamefont{Ackerstaff}}
  \bibnamefont{et~al.} (\bibinfo{collaboration}{{\sc Hermes} Collaboration}),
  \bibinfo{journal}{Nucl. Instrum. Meth.} \textbf{\bibinfo{volume}{A417}},
  \bibinfo{pages}{230} (\bibinfo{year}{1998}).

\bibitem[{\citenamefont{Nass et~al.}(2003)}]{21}
\bibinfo{author}{\bibfnamefont{A.}~\bibnamefont{Nass}} \bibnamefont{et~al.},
  \bibinfo{journal}{Nucl. Instrum. Meth.} \textbf{\bibinfo{volume}{A505}},
  \bibinfo{pages}{633} (\bibinfo{year}{2003}).

\bibitem[{\citenamefont{Airapetian et~al.}(2005)}]{22}
\bibinfo{author}{\bibfnamefont{A.}~\bibnamefont{Airapetian}}
  \bibnamefont{et~al.} (\bibinfo{collaboration}{{\sc Hermes} Collaboration}),
  \bibinfo{journal}{Nucl. Instrum. Meth.} \textbf{\bibinfo{volume}{A540}},
  \bibinfo{pages}{68} (\bibinfo{year}{2005}).

\bibitem[{\citenamefont{Ingelman et~al.}(1997)\citenamefont{Ingelman, Edin, and
  Rathsman}}]{lep}
\bibinfo{author}{\bibfnamefont{G.}~\bibnamefont{Ingelman}},
  \bibinfo{author}{\bibfnamefont{A.}~\bibnamefont{Edin}}, \bibnamefont{and}
  \bibinfo{author}{\bibfnamefont{J.}~\bibnamefont{Rathsman}},
  \bibinfo{journal}{Comp. Phys. Commun.} \textbf{\bibinfo{volume}{101}},
  \bibinfo{pages}{108} (\bibinfo{year}{1997}).

\bibitem[{\citenamefont{Akushevich et~al.}(1999)\citenamefont{Akushevich,
  Boettcher, and Ryckbosch}}]{rad}
\bibinfo{author}{\bibfnamefont{I.}~\bibnamefont{Akushevich}},
  \bibinfo{author}{\bibfnamefont{H.}~\bibnamefont{Boettcher}},
  \bibnamefont{and} \bibinfo{author}{\bibfnamefont{D.}~\bibnamefont{Ryckbosch}}
  (\bibinfo{year}{1999}), \eprint{hep-ph/9906408}.

\bibitem[{\citenamefont{Brun et~al.}(1978)\citenamefont{Brun, Hagelberg,
  Hansroul, and Lassalle}}]{gea}
\bibinfo{author}{\bibfnamefont{R.}~\bibnamefont{Brun}},
  \bibinfo{author}{\bibfnamefont{R.}~\bibnamefont{Hagelberg}},
  \bibinfo{author}{\bibfnamefont{M.}~\bibnamefont{Hansroul}}, \bibnamefont{and}
  \bibinfo{author}{\bibfnamefont{J.}~\bibnamefont{Lassalle}},
  \bibinfo{journal}{{\sc Cern} Report {\sc Cern}-DD-78-2-REV}
  (\bibinfo{year}{1978}).

\bibitem[{\citenamefont{Mo and Tsai}(1969)}]{MoTsai}
\bibinfo{author}{\bibfnamefont{L.~W.} \bibnamefont{Mo}} \bibnamefont{and}
  \bibinfo{author}{\bibfnamefont{Y.~S.} \bibnamefont{Tsai}},
  \bibinfo{journal}{Rev. Mod. Phys.} \textbf{\bibinfo{volume}{41}},
  \bibinfo{pages}{205} (\bibinfo{year}{1969}).

\end{thebibliography}
\end{document}